\documentclass[epj]{svjour}
\usepackage{graphics}
\usepackage{xcolor}
\usepackage{soul}
\setstcolor{blue}
\usepackage{booktabs} 

\begin{document}
\title{Nuclear $\beta$-decay half-lives within the subtracted second random-phase approximation}

\author{D. Gambacurta \inst{1} \and M. Grasso\inst{2}}                     
\offprints{}          
\institute{INFN-LNS, Laboratori Nazionali del Sud, 95123 Catania, Italy \and Universit\'e Paris-Saclay, CNRS/IN2P3, IJCLab, 914Concerning5 Orsay, France}
\date{Received: date / Revised version: date}
%
\abstract{
We employ, within the framework of Skyrme energy-density functional theory, the subtracted second random-phase approximation, recently developed for charge-exchange excitations, to compute $\beta$-decay half-lives in four nuclei, $^{24}$O,  $^{34}$Si, $^{78}$Ni, and $^{132}$Sn. Following our recent results on the description of the Gamow-Teller strength, we proceed coherently in the present work by computing $\beta$-decay half-lives using the bare value of the axial-vector coupling constant $g_A$. Half-lives are thus obtained, within the allowed Gamow-Teller approximation, without the use of any \textit{ad hoc} quenching factors. A genuine quenching is indeed  microscopically introduced  in our model owing to the  correlations induced by the  coupling of one-particle one-hole  configurations with two-particle two-hole ones.  The role of the so-called $J^2$ terms is also studied. By comparing our results with experimental data, we show a general improvement of  $\beta$-decay half-lives with respect to results obtained within the commonly used Random Phase Approximation (RPA). The inclusion of the two-particle two-hole configurations produces a more fragmented and richer spectrum within the $\beta$-window, resulting in lower $\beta$ half-lives with respect to the RPA ones.\\
\PACS{
      {PACS-key}{discribing text of that key}   \and
      {PACS-key}{discribing text of that key}
     } 
} 
\maketitle

\section{Introduction}
Spin-isospin excitations, see for example Ref. \cite{Hara}, offer a powerful tool to delve into the properties  of the nuclear interaction channels in the spin-isospin decomposition. By probing these excitations, we can gain valuable insights into the fundamental nature of the nuclear force and derive stringent constraints on the corresponding strength couplings. Charge-exchange excitations \cite{Oste,Ichi}, particularly the Gamow-Teller (GT) resonance, are connected to weak interaction processes such as electron capture and $\beta$-decay. These processes play a pivotal role in various astrophysical phenomena, including stellar evolution, supernova explosions, and the formation of heavy elements \cite{langanke1,langanke2}.
 Theoretical studies on the $\beta$-decay and GT excitations  have been performed within the Shell Model approach \cite{Caurier99,Yoshida2Concerning18,Saxena2Concerning18} and  non-relativistic \cite{Engel1999,Bender2Concerning02,Fracasso2005,Fracasso2007,Bai2013,Martini2014,Bai2014,Sarriguren2018,Deloncle2017}  and relativistic \cite{Paar2004,Niu2017,Liang2008,Ravlic2024,Robin2024}
approaches based on the Random Phase Approximation (RPA) or Quasi-particle RPA (QRPA) methods.	The role of beyond mean-field correlations, which are not taken into account in RPA, in describing $\beta$-decay half-lives has been explored in various theoretical works, both based on non-relativistic \cite{niu2012,Mustonen2014,niu2015,niu2016,niu2018,Engel2024} and relativistic \cite{Marketin2012,Litvinova2014,robin2016,robin2018,robin2019} approaches.

The impact of beyond mean-field correlations can be also  investigated by using the Subtracted Second RPA (SSRPA), a sophisticated theoretical framework that was first introduced in Ref. \cite{gamba2015} and applied in several studies for charge-conserving  excitations \cite{gamba2016,gamba2018a,gamba2019,gamba2020a}.  More recently, the SSRPA has been extended for the study of charge-exchange (CE) transitions \cite{gamba2020}, and a comprehensive analysis of GT resonances in several isotopes has been performed \cite{gamba2020,gamba2022}.  The effect of the tensor force within the SSRPA has been also recently studied in the case of the GT excitations and $\beta$-decay \cite{Sagawa22,Sagawa23}. The subtraction procedure \cite{tselyaev} is of crucial importance when the Second RPA (see for example Ref. \cite{Speth} and references there in), is applied within the Energy Density Functional theory, in order to overcome double-counting problems and instabilities due to the use of effective interactions. This issue was solved in past Second RPA calculations of the GT strength by neglecting the real part of the self-energy correction induced by the coupling with the two-particle two-hole ($2p-2h$) configurations, see for example Ref. \cite{Brink}.

The total GT strength of the two distinct branches satisfies the Ikeda sum rule \cite{Ikeda}, equaling 3(N-Z), where N and Z represent the neutron and proton numbers, respectively. This sum rule, being model-independent, provides a robust benchmark for theoretical models. However, while theoretical calculations often preserve the total sum rule, they frequently overestimate the experimental strength observed within a specific energy range. This discrepancy, often referred to as the "missing strength problem," is a well-known limitation of theoretical models, attributed to missing correlations such as those arising from $2p-2h$ excitations \cite{Brink,bh} or two-body currents \cite{twobody1,twobody2}. The so-called quenching factor is then  introduced in theoretical models to reconcile the discrepancy between theoretical and experimental results. 
Moreover, as the GT strength is  spread out over a broad energy range, some of it lying in the continuum, the experimental measurement is very difficult, making even harder the comparison with theory. In recent studies \cite{gamba2020,gamba2022}, we have demonstrated that explicitly incorporating $2p-2h$ configurations within the SSRPA framework, in addition to the one-particle one-hole ($1p-1h$) excitations considered in the RPA, significantly reduces the needed quenching factors, in some cases avoiding them. For instance, in the case of $^{48}$Ca, the cumulative strength calculated with SSRPA exhibits excellent agreement with experimental measurements across the entire energy range, without the need for any \textit{ad hoc} quenching factor \cite{gamba2020}. While the agreement between SSRPA and experimental data is substantially improved for heavier nuclei \cite{gamba2022} compared to RPA, a slight residual quenching (10-15\%) may still persist. This residual quenching could potentially be attributed to other effects not explicitly included in the model, such as short-range correlations and non-nucleonic degrees of freedom.

It's important to note that within RPA calculations, quenching factors are typically introduced to match experimental strength distributions measured up to higher energies, often 15-20 MeV or more, depending on the nucleus. For example, Ref. \cite{Sagawa19} presents RPA quenching factors derived for various nuclei. However, in the context of $\beta$-decay half-lives, which are primarily sensitive to the low-energy strength (see Eq. (\ref{niu1})), the primary challenge of RPA lies in its underestimation of this strength, leading to overestimated or even infinite half-lives. As will be shown in the next section, due to the limited configuration space (e.g., only $1p-1h$ type), the RPA generally underestimates the strength distribution within the $\beta$-decay window. This results in an overestimation of half-lives, which can, in some cases, be predicted as infinite, as also shown in other RPA calculations \cite{niu2018,Sagawa23}).

A reliable theoretical model should be capable of accurately describing both the low-energy and full-energy strength distributions without resorting to any experimentally tuned quenching factors. Our recent work demonstrates that the explicit inclusion of $2p-2h$ configurations within the SSRPA framework represents a significant step forward in this direction. Moreover, by addressing the limitations of traditional models and providing a more consistent and accurate framework for nuclear weak processes, this study contributes to a deeper understanding of fundamental nuclear physics and its implications for astrophysics and particle physics.  For example,  the present work is also expected to have direct implications for the computation of nuclear matrix elements that are essential for predicting neutrino-less double $\beta$-decay half-lives. 	As a matter of fact, a model capable of reliably reproducing both GT excitations and single $\beta$-decay half-lives without the need for empirical adjustments is a strong candidate for providing unambiguous predictions of neutrino-less double $\beta$-decay matrix elements using unquenched quantities.
However, a reliable and predictive description of the $\beta$-decay processes requires also the inclusion of other important ingredients. For example, in the case of neutron-rich nuclei, forbidden transitions can play an important role \cite{Robin2024,Marketin2016}. Also, the inclusion of pairing correlations would be needed in the SSRPA, to treat open-shell systems. Work in this direction is underway. Finally, nuclear deformation can also play an important role, both in extending $\beta$-decay studies across the nuclear chart and, more specifically, in the case of neutrino-less double $\beta$-decay, for which the main candidate nuclei are indeed non-spherical. This would require developing the SSRPA within a deformed basis, which involves the breaking of the rotational invariance. Consequently, the computational cost would be significantly higher than in the spherical case, where the j-coupled scheme drastically reduces the dimensions of the SSRPA matrix. Therefore, unless resorting to approximations, such as the diagonal approximation \cite{gamba2010} or targeting specific narrow regions of the energy spectrum, this extension is unlikely to be implemented in the near future for systematic studies.

\section{The SSRPA equations}
The Second RPA formalism and the subtraction procedure are described in detail in Refs \cite{gamba2015,gamba2016,gamba2010}. A brief review of the main aspects follows. In the RPA approximation the excitations operators are assumed to be a linear superposition
of $1p-1h$ operators:
\begin{equation}\label{rpa_op}
  Q_{\nu}^{\dag}=\sum_{ph}X_{ph}^{\nu}a_{p}^{\dag}a_{h}-\sum_{ph}Y_{ph}^{\nu}a_{h}^{\dag}a_{p},
\end{equation}
where for notation simplicity, the coupling to total quantum numbers is not indicated.

In the case of the SRPA, the $2p-2h$ configurations are explicitly considered in the description of the
excitations operators, having the following structure:
\begin{displaymath}\label{srpa_op}
    Q_{\nu}^{\dagger}=\sum_{ph}X_{ph}^{\nu}a_{p}^{\dag}a_{h}-\sum_{ph}Y_{ph}^{\nu}a_{h}^{\dag}a_{p}
\end{displaymath}
\begin{equation}
+    \sum_{p<p',h<h'}(X_{php'h'}^{\nu}
a_{p'}^{\dag}a_{p}^{\dag}a_{h'}a_{h}-Y_{php'h'}^{\nu}a_{h}^{\dag}
a_{h'}^{\dag}a_{p}a_{p'}).
\end{equation}

In both cases, the energies $\omega_{\nu}$ of the excited states and their wave function ( e.g. the $X's$ and $Y's$ amplitudes)
are  obtained by solving an eigenvalue problem of this form:
\begin{equation}\label{eq_srpa}
\left(\begin{array}{cc}
  \mathcal{A} & \mathcal{B} \\
  -\mathcal{B}^{*} & -\mathcal{A}^{*} \\
\end{array}\right)
\left(%
\begin{array}{c}
  \mathcal{X}^{\nu} \\
  \mathcal{Y}^{\nu} \\
\end{array}%
\right)=\omega_{\nu}
\left(%
\begin{array}{c}
  \mathcal{X}^{\nu} \\
  \mathcal{Y}^{\nu} \\
\end{array}%
\right).
\end{equation}
In the RPA case  the $\mathcal{A}$ and $\mathcal{B}$ matrices describe the coupling among the $1p-1h$ configurations,
while in SSRPA, more general block matrices appear, describing the coupling of the $1p-1h$ configurations with
the $2p-2h$ configurations and of the $2p-2h$ configurations among themselves. For example, the $\mathcal{A}$ matrix can be written as a block of matrices,

\begin{displaymath}
\mathcal{A}=\left(\begin{array}{cc}
  A_{11'} & A_{12} \\
  A_{21} & A_{22'} \\
\end{array}\right),
\end{displaymath}
where '1' and '2' stand for $1p-1h$ and $2p-2h$ configurations, respectively.

The SRPA model, when applied within the Energy Density Functional framework suffers of double-counting issues and potential instabilities that are removed by employing the so-called subtraction procedure \cite{gamba2015,tselyaev} that corrects the response to make it consistent with ground-state density-functional
theory at zero frequency. The subtraction procedure consists in subtracting in the $A_{11}$ block
of the SRPA matrix the quantity

\begin{equation}
E_{11^{\prime}}  =  -\sum_{2,2^{\prime}} A_{12} (  A_{22^{\prime}})^{-1}
A_{2^{\prime}1^{\prime}}.
\label{eq:correction}
\end{equation}

\section{Results}
In this work, we adopt a consistent approach to calculate $\beta$-decay half-lives using the SSRPA model, without the use of  any renormalization factor, in line with our previous work on total GT strength calculations \cite{gamba2020,gamba2022}. The Skyrme interaction \cite{Vautherin} is employed consistently in the description of the ground state, within the Hartree-Fock approximation, and of the excited states, both at RPA and SSRPA level. In order to study the sensitivity of the results, different Skyrme parametrizations are used, as discussed in more detail below. Given that GT strengths are computed using the bare GT operator, we employ the bare value of the axial-vector coupling constant $g_A=1.26$, in our half-life calculations. This contrasts with the common practice of using a quenched value of $g_A=1$ in the literature. The quenching factor is often introduced to account for missing correlations. 


\begin{figure}
	\resizebox{0.5\textwidth}{!}{\includegraphics{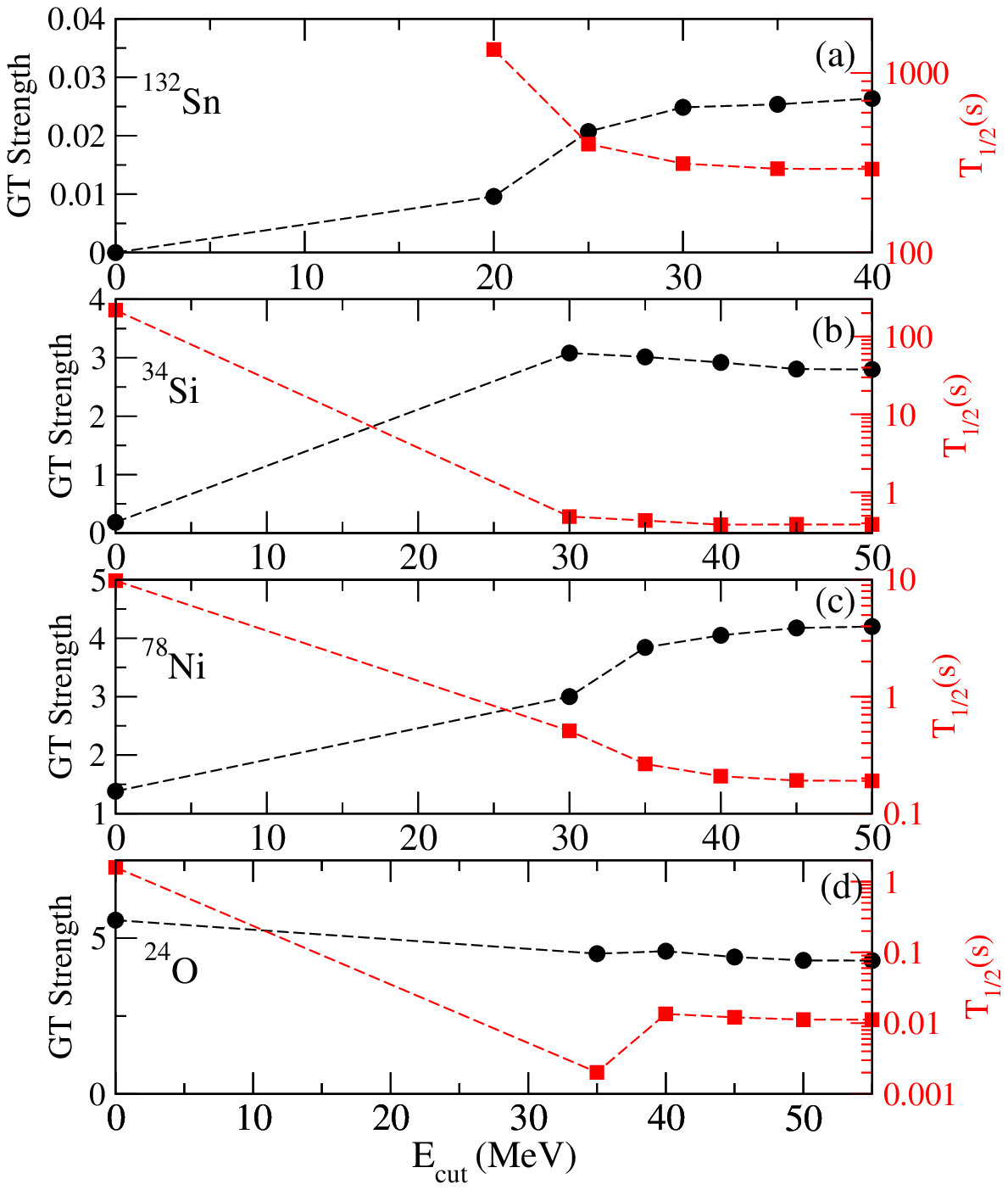}}
	\caption{SSRPA integrated GT strength in the $\beta$-window (black circle, right y-axis) and corresponding $\beta$-decay half-lives  (red square, left y-axis) for $^{132}$Sn (panel (a)),  $^{34}$Si (panel (b)),  and $^{78}$Ni (panel (c)) and $^{24}$O (panel (d))  as a function of the energy cutoff $E_{cut}$ on the $2p-2h$ configurations for the SGII interaction. The values for $E_{cut}=0$ MeV, corresponding to the RPA results, are shown as a reference. For $^{132}$Sn the RPA half-life is infinite as the total strength is zero. }
	\label{Fig:Fig0}
	\end{figure}

\begin{figure}
\resizebox{0.5\textwidth}{!}{\includegraphics{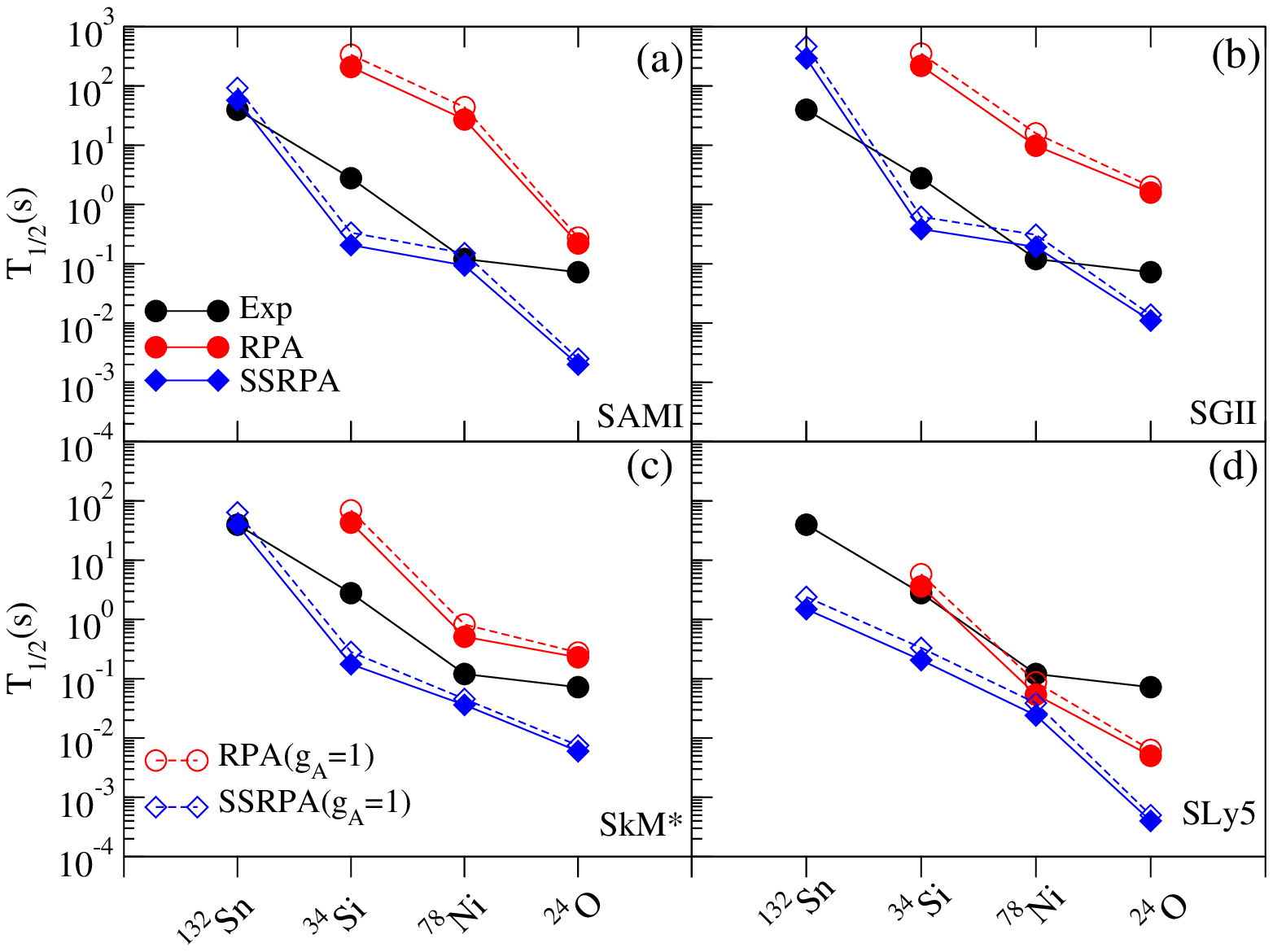}}
\caption{Experimental $\beta$-decay half-lives for $^{132}$Sn, $^{24}$O, $^{34}$Si \cite{exp} and $^{78}$Ni \cite{xu}, compared with those obtained within the RPA and SSRPA models. Full (empty) symbols are obtained by using the bare $g_A=1.26$ ($g_A=1.00$) constant. The RPA values that are not shown correspond to infinite half-life values. See the text for more details.}
\label{Fig:Fig1}
\end{figure}

We apply the SSRPA model in the study of $\beta$-decay half-lives for four representative nuclei: $^{24}$O, $^{34}$Si,  $^{78}$Ni, and $^{132}$Sn. These nuclei can be safely considered as spherical non-superfluid nuclei where pairing correlations are not expected to play an important role. In the case of $^{34}$Si, it has been shown that pairing correlations can be neglected \cite{Grasso2009}. The SSRPA results are  compared to those obtained within  the conventional RPA approach.

\begin{figure}
\resizebox{0.5\textwidth}{!}{\includegraphics{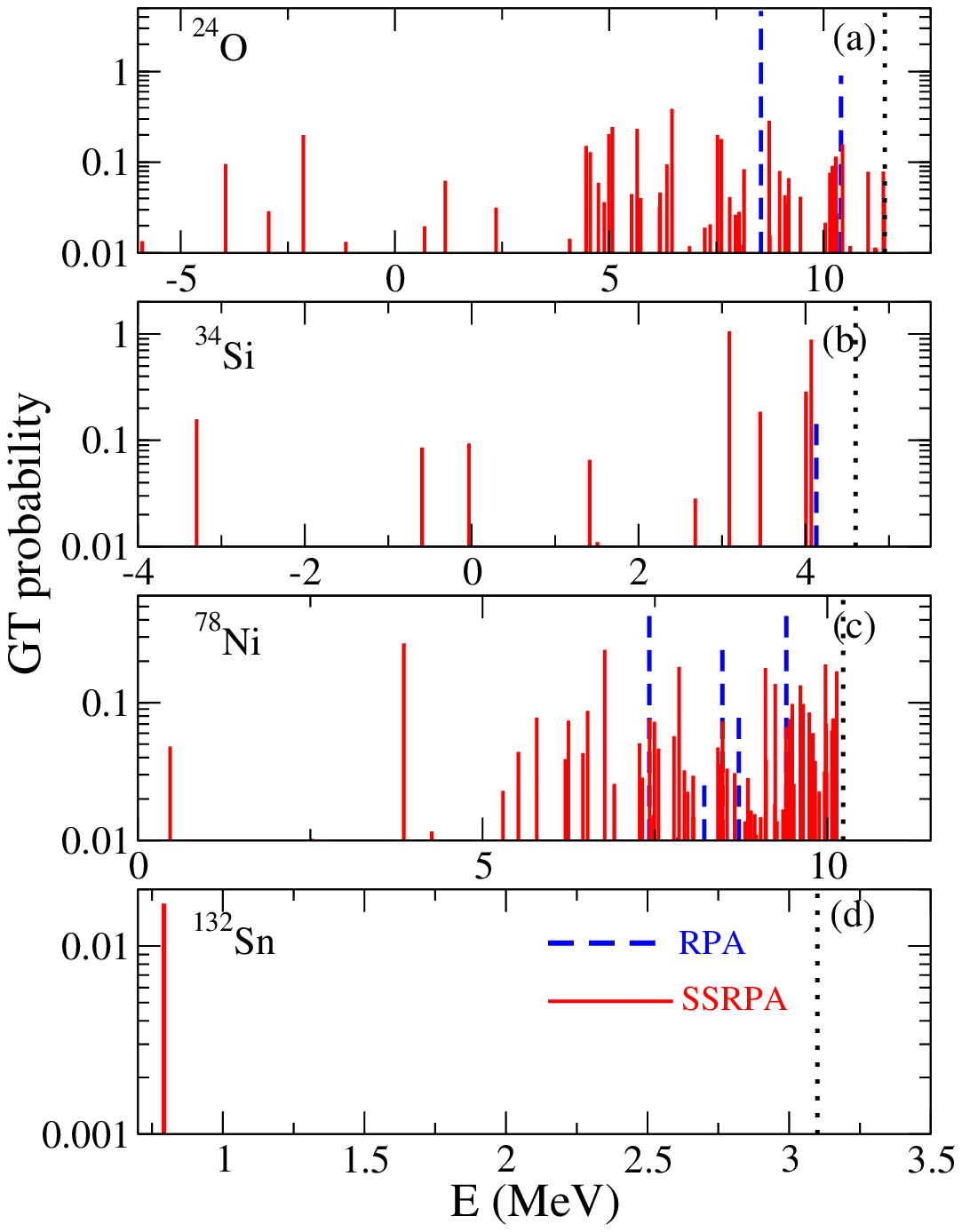}}
\caption{The GT strength distributions with respect to excitation energies referred to the ground state of daughter nucleus,  in $^{24}$O, $^{34}$Si, $^{78}$Ni, and $^{132}$Sn with SGII obtained in RPA (blue-dashed line) and SSRPA (red-solid line).  The vertical dotted lines show the experimental value of  $Q_{\beta}$ value. See the text for more details.}
\label{Fig:SGII-Strength}
\end{figure}

If $E$ is the excitation energy referred to the ground state of the daughter nucleus, the half-life is given, in the allowed GT approximation \cite{engel,niu}, by the following expression:
\begin{equation}
T_{1/2}=\frac{D}{g_A^2 \int_0^{Q_{\beta}}S(E) f(Z,\omega) dE},
\label{niu1}
\end{equation}
where
\begin{equation}
D=6163.4 \pm 3.8 \; s
\label{DD}
\end{equation}
and
$Q_{\beta}$ is the $Q$ value: $Q_{\beta}= \Delta_{nH} - \Delta B$. $\Delta B $ is the difference of the binding energies of the mother and the daughter nuclei, $B_{mother}-B_{daughter}$, and $\Delta_{nH} = 0.78227 $ MeV is the mass difference between the neutron and the hydrogen. In the present work, $\Delta B$ is evaluated by using the experimental 	binding energy difference. In Eq. (\ref{niu1}),
$g_A$ is the axial-vector coupling constant of the weak interaction. $S(E)$ is the GT strength function and
$f(Z,\omega)$ is the integrated phase volume or phase--space volume (which contains the lepton kinematics) \cite{niu}.
The GT strength function is calculated by using the one-body transition operator
\begin{equation}
\hat{O}^{-}=\sum_{i=1}^{A} \sum_{\mu} \sigma_{\mu}(i) \tau^{-}(i),
\label{oper}
\end{equation}
where  $\tau^{-}$ is the isospin lowering  operator, $\tau^{-}=t_x - it_y$, $\sigma_{\mu}$ is the spin operator, and  $A$ is the number of nucleons.

Recently, the self-consistent SSRPA has been applied to compute the $\beta$-decay half-life of $^{78}$Ni \cite{gamba2020}. Using the Skyrme interaction SGII, we found excellent agreement with the experimental value reported in Ref. \cite{hosmer}. Subsequent experimental measurements have refined the half-life of $^{78}$Ni to $T_{1/2}=122.2 \pm 5.1$ ms \cite{xu}. Importantly, this updated value does not alter the conclusion drawn in Ref. \cite{gamba2020}: the Skyrme interaction SGII continues to provide a highly accurate prediction.
In the present work, we will reference this more recent experimental value. The SSRPA framework has also been applied in a recent study \cite{Sagawa23}, corroborating the findings of Ref. \cite{gamba2020} and additionally exploring the impact of the tensor force on $\beta$-decay. In  Ref. \cite{Sagawa23}, besides the case of  $^{34}$Si,  $^{78}$Ni, and $^{132}$Sn also the nucleus $^{68}$Ni was studied, showing in general a more problematic behavior in reproducing the experimental data, also within the SSRPA. In our calculations, a very slow convergence of the SSRPA results for this nucleus with respect to the cutoff on the $2p-2h$  configurations was indeed found. We also found a even stronger dependence of these results on the $J^2$-term (see discussion later), especially in the SGII case. However, since $^{68}$Ni is indeed super-fluid in the neutron channel, the inclusion of pairing correlations would be indeed needed. For these reasons, in the present work, we do not show the result on  $^{68}$Ni as done in \cite{Sagawa23}.

\begin{figure}
\resizebox{0.5\textwidth}{!}{\includegraphics{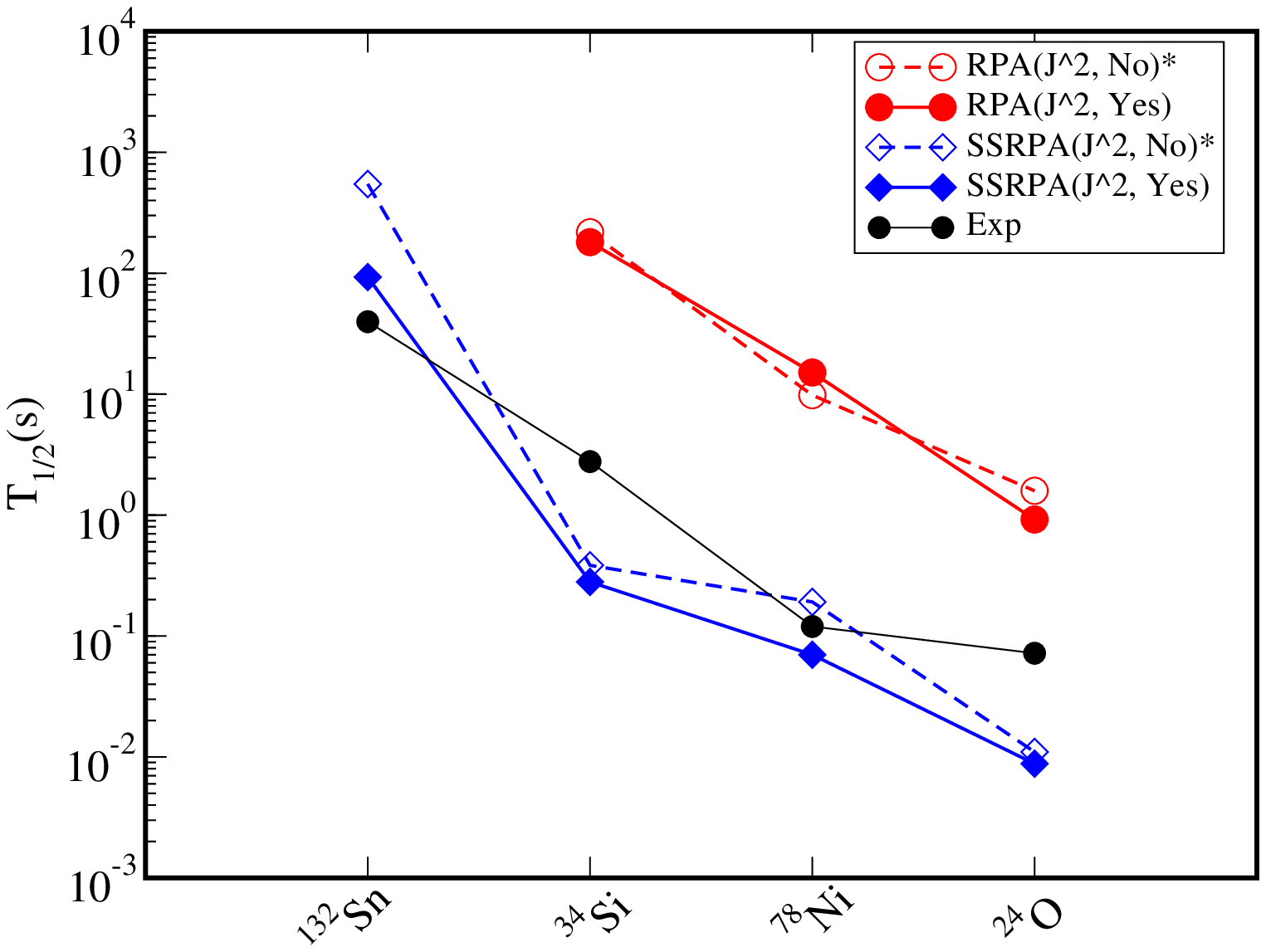}}
\caption{Comparison between the RPA and SSRPA with and without the $J^2$ terms. The SGII interaction is used. The ``*'' symbol in the legend indicates  that the SGII force has been derived without $J^2$ terms.}
\label{Fig:Fig2-SGII}
\end{figure}

\begin{figure}
\resizebox{0.5\textwidth}{!}{\includegraphics{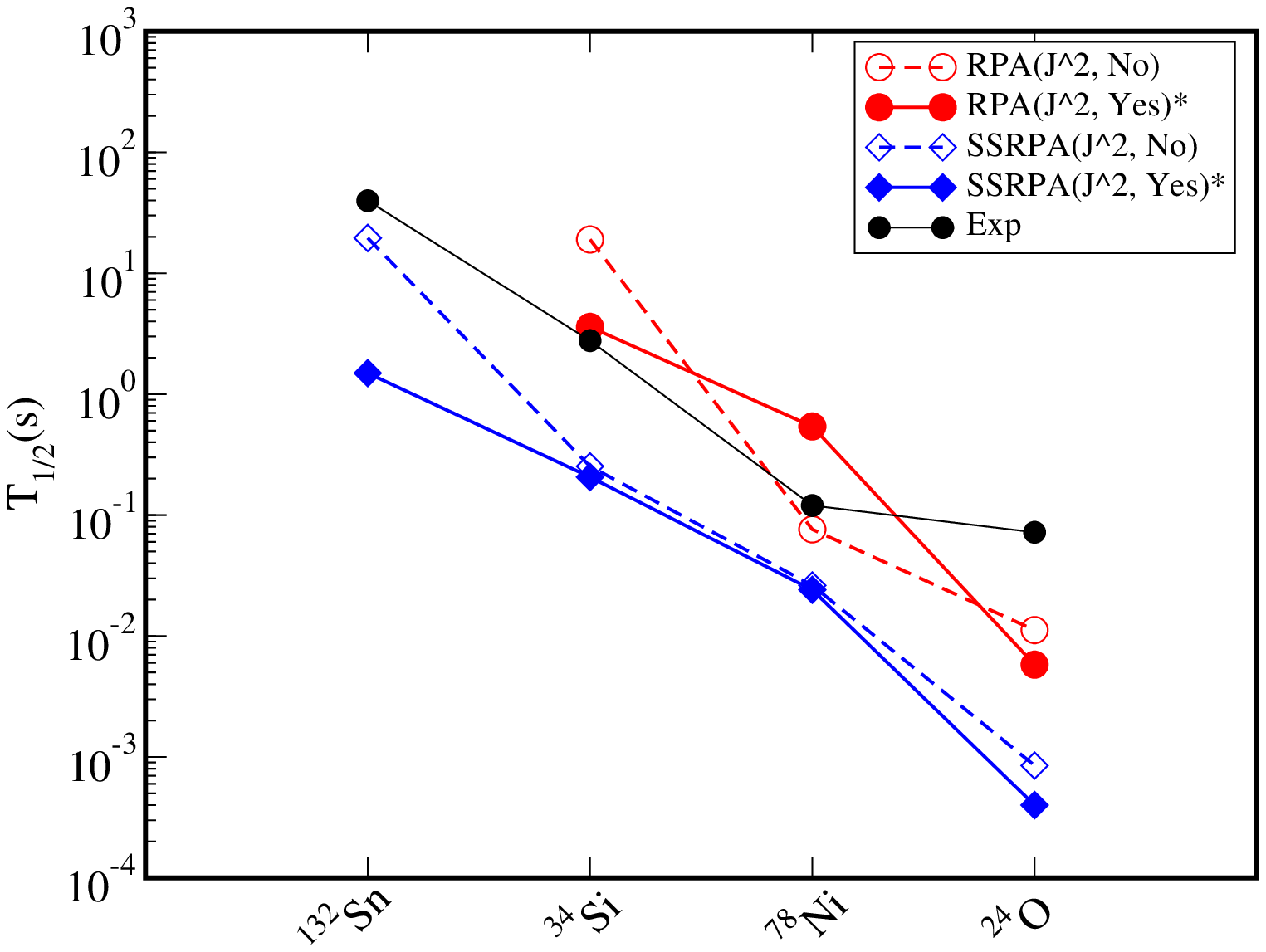}}
\caption{Comparison between the RPA and SSRPA with and without the $J^2$ terms. The SLy5 interaction is used. The ``*'' symbol in the legend  indicates that the SLy5 force has been derived with $J^2$ terms.}
\label{Fig:Fig3-SLY5}
\end{figure}

Calculations are performed by using energy cutoffs on the $1p-1h$ and $2p-2h$ are chosen for each nucleus and Skyrme force such as to obtain stable results. More precisely, all the $1p-1h$ configurations with an unperturbed energy less than 100 MeV are included, exhausting the Ikeda sum rule \cite{Ikeda} within less than 1\% for all the considered cases. The convergence with respect to the energy cutoff $E_{cut}$ in the $2p-2h$ space has been checked by performing, for each nucleus and interaction, the SSRPA calculations with progressively increasing $E_{cut}$, incrementing it by by 5 MeV at each step. This energy step ensures a significant increase in the $2p-2h$ configuration space for all the studied nuclei. The calculation is considered convergent when the relative difference between two successive evaluations of the $\beta$-decay half-life is less than 1\%. The specific energy cutoff at which convergence is achieved may exhibit a slight dependence on the nucleus and the employed interaction. However, the general trend indicates that stability is reached by employing the following values for $E_{cut}$: 50-55 MeV (for $^{24}$O), 40-45 MeV (for $^{34}$Si and $^{78}$Ni), and 30-35 MeV (for $^{132}$Sn).  As an illustrative example, we consider SGII interaction \cite{SGII}, which exhibits slower convergence, necessitating higher cutoff values than other parametrizations to achieve stability. In Figure \ref{Fig:Fig0}, the SSRPA integrated GT strength within the $\beta$-window (indicated by black circles on the right y-axis) and the corresponding $\beta$-decay half-lives (shown as red squares on the left y-axis) are plotted as a function of the energy cutoff $E_{cut}$  for $^{132}$Sn (panel (a)), $^{34}$Si (panel (b)), $^{78}$Ni (panel (c)), and $^{24}$O (panel (d)). The values for $E_{cut} = 0$ MeV, corresponding to the RPA case, are presented as a reference. One can observe that, in all cases, the $\beta$-decay half-life   decreases from RPA to SSRPA, until the convergence is reached by increasing the energy cutoff $E_{cut}$.  Concerning the integrated strength, one can see that the SSRPA provides larger values than the RPA, with the exception of the $^{24}$O case, where we observe a slight decrease when moving from the RPA to SSRPA (see also panel (a) of Fig. \ref{Fig:SGII-Strength}). Nevertheless, even for the $^{24}$O, the SSRPA half-life is lower than the RPA, because of the  phase-space volume $f(Z,\omega$) appearing in Eq. (\ref{niu1}), which weights the strength of each state in an energy-dependent way. This result shows that the $\beta$-decay half-life value is determined by the strength distributions inside the $\beta$ window and not only by the integrated strength.  For $^{132}$Sn, the RPA half-life is formally infinite due to a zero total GT strength within the relevant energy window.
	
In Fig. \ref{Fig:Fig1}, the $\beta$-decay half-lives calculated within the RPA and SSRPA frameworks are compared with experimental values \cite{xu,exp}. Four Skyrme parametrizations  are here employed, namely SAMI \cite{SAMI}
(panel a), SGII \cite{SGII} (panel b), SkM$^*$ \cite{SKMS} (panel c) and SLy5 \cite{SLY4} (panel d). SSRPA results are obtained by using the diagonal approximation in the subtraction procedure, which we have verified to be reliable for this study, as discussed in Ref. \cite{gamba2022}
  Notably, none of the adopted force is able to provide a finite $\beta$ half-life for $^{132}$Sn.    As a general feature, one can see that the RPA strongly overestimate the data, with the exception of the SLy5 force for $^{34}$Si and $^{78}$Ni. On the contrary, for all the Skyrme forces and nuclei, the SSRPA present excited states within the $\beta$ window, and, as a consequence, finite half-lives are provided. As a general trend, due to the inclusion of the $2p-2h$ configurations which produces a richer and more fragmented spectrum with respect to the RPA distribution, the SSRPA half-lives are much lower, providing an overall better agreement with the experimental data. Solid and dashed symbols represent results obtained using axial-vector coupling constants of $g_A=1.26$ and $g_A=1.00$, respectively. This comparison demonstrates a genuine quenching effect within the SSRPA framework. By "genuine" quenching, we emphasize that, unlike the \textit{ad hoc} quenching factor typically employed to reproduce experimental data, this reduction  arises microscopically from the inherent coupling with $2p-2h$ configurations within the SSRPA. Furthermore, this microscopically derived quenching effect is significantly stronger than the artificial quenching achieved by simply using an effective (reduced) $g_A=1.00$ value. 
  These results are in agreement with those found in Ref. \cite{Sagawa23}, minor differences might be due to numerical reasons. In both SSRPA calculations, the inclusion of the $2p-2h$ configurations lowers the $\beta$ half-lives, generally improving the agreement with the experimental values compared to  the RPA.
The general trend of the SSRPA results is also in agreement with the results obtained within the PVC model \cite{niu2015} where the effect of more complex configurations was introduced through the coupling with collective phonons, which in the case of $^{68}$Ni seems to be more efficient than what is found in Ref. \cite{Sagawa23}.
It is also worth mentioning that, with the exception of the $^{132}$Sn case, for which any parametrization provides an infinite half-life at RPA level, the SLy5 force provides RPA half-lives in good agreement with data, in particular for $^{34}$Si and $^{78}$Ni. This force was designed to improve the Skyrme functional description of the isospin degrees of freedom, especially in nuclei far from stability. This is achieved also by fitting to properties of asymmetric nuclear matter as pseudo-data. This protocol likely explains the good agreement shown by the SLy5 force. However, including correlations beyond RPA, as done in SSRPA (see also Ref. \cite{Sagawa23}) and PVC calculations (Ref. \cite{niu2018}), generally worsens agreement with data for this parametrization. Therefore, beyond-RPA approaches using mean-field-optimized effective interactions demand careful validation. While subtraction methods address double-counting and restore the overall response, reproducing fine spectral details, like strength in narrow energy regions, as in the present case, might remain force-dependent.

In Fig. \ref{Fig:SGII-Strength}, as an illustrative case, we plot the GT strength distributions for  $^{24}$O, $^{34}$Si, $^{78}$Ni, and $^{132}$Sn. The SGII interaction is used both in the  RPA (blue-dashed lines) and SSRPA (red-solid lines) results. The spectra are shown up to the $Q_\beta$ value which is indicated by a dotted-black lines in the figure. As expected, the RPA tends to produce less fragmented distributions, with fewer peaks and a more concentrated strength. As a consequence, the RPA tends to underestimate the GT strength, as it can be seen from Fig. \ref{Fig:Fig1}, where the RPA lifetimes are systematically overestimating the experimental value. In the $^{132}$Sn case, as mentioned above, no state is provided within the RPA.
The SSRPA, on the other hand, tends to give a more realistic description of the GT strength distribution. It predicts a more fragmented distribution  as a consequence of the coupling with the $2p-2h$ configurations. The GT strength distribution is sensitive to the nuclear structure. For example, $^{24}$O and $^{78}$Ni, being neutron-rich nuclei, show a much richer strength distribution already at RPA level, which is further increasing in SSRPA.

Given the high sensitivity of half-lives to the low-energy strength distribution, it is worth  investigating the role of the so-called $J^2$-terms
\cite{Vautherin}. The latter are introduced by the central part of the interaction and contribute to the ground state energy  with terms that are proportional to the square of the current density $J$ and are not systematically taken into account in all the parametrizations of the Skyrme forces. Though their contribution to the binding energy is often negligible, they might affect the low-lying distribution of the spectra competing with those induced by the tensor force \cite{Bender}. In order to quantify their effect on the $\beta$-decay half life,  we select the SGII interaction, which has been fitted without the  $J^2$-terms and therefore  these terms are not included in the results shown in Fig. \ref{Fig:Fig1}, and the SLy5 force which on the contrary does include them. For these two parametrizations, we study the effect of switching on and off the $J^2$-terms on the $\beta$-decay half-lives, both at RPA and SSRPA level. In Fig. \ref{Fig:Fig2-SGII}, we show the results for the SGII case and one can see that the effect of the $J^2$-terms is quite negligible at RPA level for $^{34}$Si, being slightly larger for $^{78}$Ni, though including them does not improve the agreement with the data in both cases. In the case of the SSRPA, one can see that for $^{132}$Sn, the  $J^2$-terms have a significant impact on the predicted half-lives, their inclusion further improving the agreement with data. The effect is instead much weaker for the other nuclei, for which the agreement is not strongly affected. In Fig. \ref{Fig:Fig3-SLY5}, we show a similar analysis but for the SLy5  force. In the RPA case, the $J^2$-terms are impacting the lifetime for $^{34}$Si and $^{78}$Ni, acting however in opposite direction, e.g. excluding them  worsens (improves) the agreement for  $^{34}$Si ($^{78}$Ni).
The opposite is happening instead within the SSRPA, where the results are almost unchanged for $^{24}$O, $^{34}$Si and $^{78}$Ni, while for  $^{132}$Sn , excluding the  $J^2$-terms provides indeed results closer to the experimental data. Our findings indicate that the $J^2$-terms might have an important role, albeit in a nucleus- and force-dependent manner, and their effect is comparable with those of the tensor interaction recently analyzed in Ref. \cite{Sagawa23}. The competition between these two terms should be carefully taken into account, especially in the derivation of new functionals aiming  at achieving an improved description of the $\beta$-decay and GT strength distribution.
\begin{figure}
\resizebox{0.5\textwidth}{!}{\includegraphics{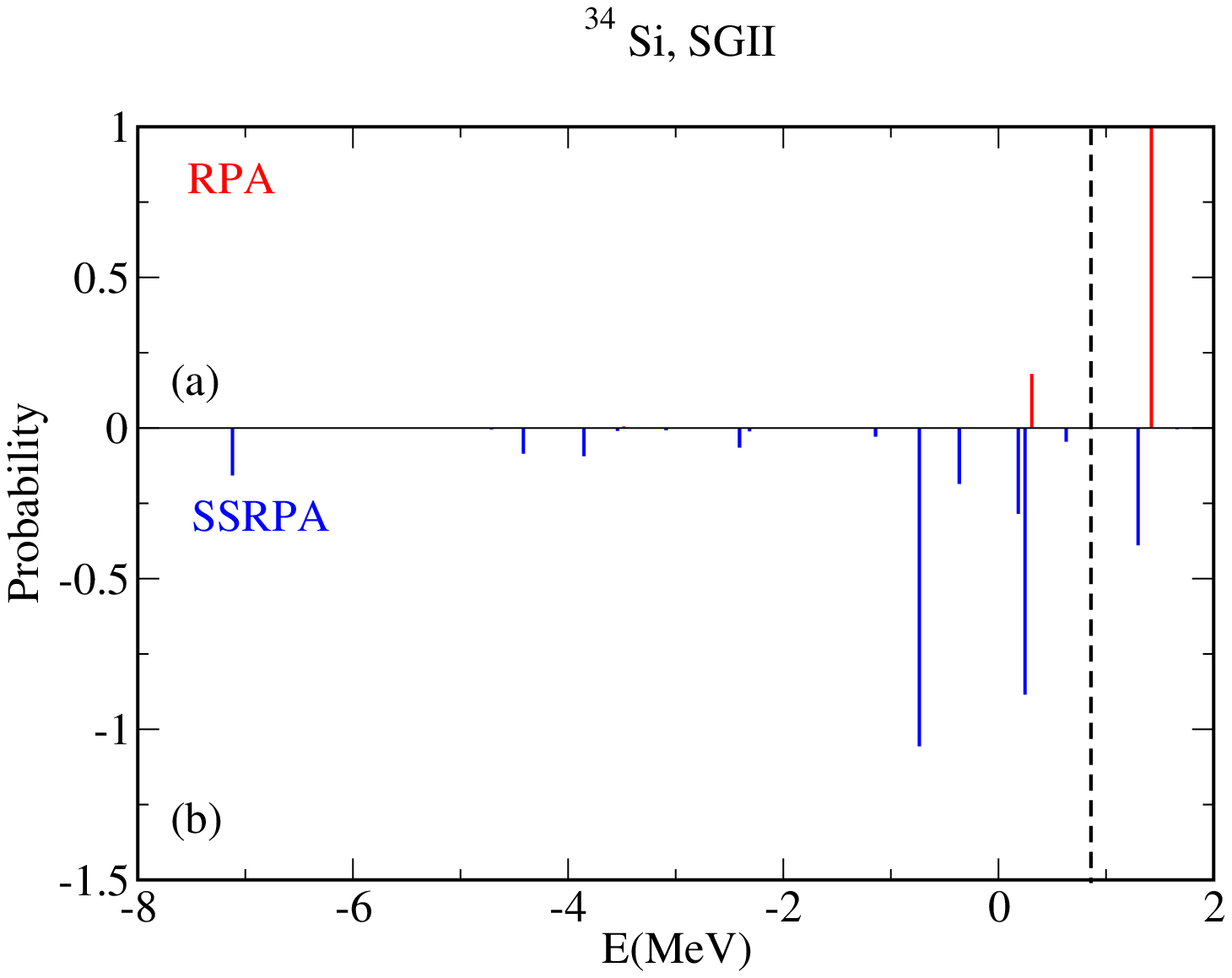}}
\caption{The GT strength distributions
with respect to the parent nucleus in $^{34}$Si with SGII obtained in RPA (upper panel)
and SSRPA (lower panel). The black dashed line represents the $\Delta_{nH}$ value.
}
\label{Fig-Si34-SGII}
\end{figure}

Finally, we  present a more detailed discussion, focusing on the  $^{34}$Si nucleus, analyzing the RPA and SSRPA wave functions of the most collective states contributing to the $\beta$ half-life. In Fig. \ref{Fig-Si34-SGII},  we show the GT strength distributions as a function of the excitation energy (with respect to the parent nucleus)  obtained in RPA (upper panel)
and SSRPA (lower panel) with the SGII interaction. For the sake of simplicity, we consider states with a transition probability larger than 0.1 and with energy lower than the $\Delta_{nH}$ value (black dashed line ), relevant to the $\beta$-decay.  In order to have a deeper insight on the properties of these states, we define:
\begin{equation}
N_{ph}^{\nu} = |X^\nu_{ph}|^2 -|Y^\nu_{ph}|^2
\label{norm}
\end{equation}
quantifying the amount of each $1p-1h$ configuration in the norm of each phonon state $\nu$, having that $\sum_{ph}N_{ph}^{\nu}=1$ in the RPA case, while in the SSRPA case, the normalizations reads as
\begin{equation}
	\sum_{ph}N_{ph}^{\nu}+	\sum_{php'h'}	N_{php'h'}^{\nu}=1
	\label{norm_tot}
\end{equation}
where
\begin{equation}
	N_{php'h'}^{\nu} = |X^\nu_{php'h'}|^2 -|Y^\nu_{php'h'}|^2
	\label{norm2}
\end{equation}
is the contribution of each $2p-2h$ configuration.

Moreover, for each excited state $\nu$, the probability to be excited $P(\nu) $, is given by the sum of each individual $1p-1h$ contribution $b_{ph}(\nu)$, that is the product of the wave-function amplitude $A_{ph}^\nu=X_{ph}^\nu - Y_{ph}^\nu$ multiplied by the matrix element of the transition operator $F_{ph}^\lambda$, having thus
\begin{displaymath}
P(\nu) = |\sum_{ph} b_{ph}(\nu)|^2 =|\sum_{ph} (X_{ph}^\nu - Y_{ph}^\nu)
 F_{ph}^\lambda|^2=
 \end{displaymath}
\begin{equation}
 |\sum_{ph} A_{ph}^\nu
 F_{ph}^\lambda|^2.
 \label{prob}
\end{equation}
\begin {table}
\caption {\label{tab-1} Particle-hole configurations which give
    the major contributions to the the norm (upper part) and to the GT transition probability (lower part) for the RPA state located at 0.310 MeV and 1.422 MeV. See upper panel of Fig. \ref{Fig-Si34-SGII}. The superscripts $\pi$, $\nu$ refer to
    proton and neutron states, respectively. See Eqs (\ref{norm}) and (\ref{prob}) for the definition of each quantity. The $E_{1p-1h}$ energies are given in  MeV units.  
    }
    \resizebox{\columnwidth}{!}{%
 \begin{tabular}{|ccccc|}
 \hline
 \hline
&&E$_{RPA}$=       0.310    MeV &&\\
 \hline
 $1p-1h$ conf. &$E_{1p-1h}$   &$N_{ph}$& &\\
$[\pi2s_{1/2},\nu2s_{1/2}]$&     -1.379&   0.440  &&\\
$[\pi1d_{3/2},\nu1d_{3/2}]$&     -0.726&   0.513  &&\\
&$N_1=1.000$&$N_2=0.000 $ &&\\
 \hline
 $1p-1h$ conf. &$E_{1p-1h}$   &$A_{ph}$& $F_{ph}$&$b_{ph}$\\
 \hline
$[\pi1d_{3/2},\nu1d_{5/2}]$&   6.038&  -0.030&  3.094&  -0.094 \\
$[\pi2s_{1/2},\nu2s_{1/2}]$&  -1.379&   0.663&  2.443&   1.620 \\
$[\pi1d_{3/2},\nu1d_{3/2}]$&  -0.726&   0.716& -1.548&  -1.109 \\
 \hline
 $\sum b_{ph}$&&  &&   0.423\\
        $P(\nu)$&&  &&   0.179\\
 \hline
 \hline
&&E$_{RPA}$=       1.422    MeV $>\Delta_{nH}$ &&\\
 \hline

 $1p-1h$ conf. &$E_{1p-1h}$   &$N_{ph}$& &\\
$[\pi1d_{3/2},\nu1d_{5/2}]$&      6.038&   0.049  &&\\
$[\pi2s_{1/2},\nu2s_{1/2}]$&     -1.379&   0.509  &&\\
$[\pi1d_{3/2},\nu2s_{1/2}]$&      2.161&   0.089  &&\\
$[\pi1d_{3/2},\nu1d_{3/2}]$&     -0.726&   0.346  &&\\
  &&&$N_1=       1.000$&$N_2=       0.000 $ \\
 \hline
 $1p-1h$ conf. &$E_{1p-1h}$   &$A_{ph}$& $F_{ph}$&$b_{ph}$\\
 \hline
$[\pi1d_{3/2},\nu1d_{5/2}]$&      6.038&      -0.221&       3.094&      -0.684 \\
$[\pi2s_{1/2},\nu2s_{1/2}]$&     -1.379&       0.714&       2.443&       1.744 \\
$[\pi1d_{3/2},\nu1d_{3/2}]$&     -0.726&      -0.588&      -1.548&       0.911 \\
 \hline
 $\sum b_{ph}$&&  &&   1.977\\
        $P(\nu)$&&  &&   3.910\\
 \hline
 \end{tabular}
 }
 \end{table}

\begin {table}
\caption {\label{tab-2} Particle-hole  configurations which give
	the major contributions to the the norm (upper part) and to the GT transition probability (lower part) for the  for the SSRPA states within the $\beta$ window  located at -7.120, -0.735, -0.364 MeV . See upper panel of Fig. \ref{Fig-Si34-SGII}. The superscripts $\pi$, $\nu$ refer to
	proton and neutron states, respectively. See Eqs (\ref{norm}) and (\ref{prob}) for the definition of each quantity.  The $E_{1p-1h}$ and $E_{2p-2h}$ energies are given in  MeV units.  
}
\resizebox{\columnwidth}{!}{%
	\begin{tabular}{ccccc}
		\toprule
		&&E$_{SSRPA}$=      -7.120    MeV &&\\
		\hline
		$1p-1h$ conf. &$E_{1p-1h}$   &$N_{ph}$& &\\
		$[\pi2s_{1/2},\nu2s_{1/2}]^{J=1^+}$&     -1.379&   0.017  &&\\
		$[\pi2s_{1/2},\nu1d_{3/2}]^{J=1^+}$&     -4.266&   0.012  &&\\
		&&&$N_1=       0.032$&$N_2=       0.968 $ \\
		\hline
		$1p-1h$ conf. &$E_{1p-1h}$   &$A_{ph}$& $F_{ph}$&$b_{ph}$\\
		\hline
		$[\pi1d_{3/2},\nu1d_{5/2}]^{J=1^+}$&      6.038&       0.026&       3.094&       0.079 \\
		$[\pi2s_{1/2},\nu2s_{1/2}]^{J=1^+}$&     -1.379&       0.132&       2.443&       0.322 \\
		\hline
		$\sum b_{ph}$&&  &&   0.397\\
		$P(\nu)$&&  &&   0.157\\
		\hline
		\multicolumn{3}{c}{$2p-2h$ conf. }&$E_{2p-2h}$   &$N_{php'h'}$\\
		\multicolumn{3}{c}{$\big[[\pi1d_{5/2}\nu1d_{3/2}]_{J_H=1}[\pi2s_{1/2}\pi2s_{1/2}]_{J_P=0}\big]_{J_{T}=1}$} &   5.867 &  0.723  \\
		\multicolumn{3}{c}{$\big[[\pi1d_{5/2}\nu1d_{3/2}]_{J_H=3}[\pi2s_{1/2}\pi1d_{3/2}]_{J_P=2}\big]_{J_{T}=1}$} &   9.407 &  0.099  \\
		\multicolumn{3}{c}{$\big[[\pi1d_{5/2}\nu1d_{3/2}]_{J_H=2}[\pi2s_{1/2}\pi1d_{3/2}]_{J_P=2}\big]_{J_{T}=1}$} &   9.407 &  0.054  \\
		\midrule 
	\end{tabular}
}
%
%
%
\resizebox{\columnwidth}{!}{%
\begin{tabular}{ccccc}
	\toprule
	&&E$_{SSRPA}$=      -0.735    MeV &&\\
	\hline
	$1p-1h$ conf. &$E_{1p-1h}$   &$N_{ph}$& &\\
	$[\pi2s_{1/2},\nu2s_{1/2}]^{J=1^+}$&     -1.379&   0.109  &&\\
	&&&$N_1=       0.125$&$N_2=       0.875 $ \\
	\hline
	$1p-1h$ conf. &$E_{1p-1h}$   &$A_{ph}$& $F_{ph}$&$b_{ph}$\\
	\hline
	$[\pi1d_{3/2},\nu1d_{5/2}]^{J=1^+}$&      6.038&       0.023&       3.094&       0.070 \\
	$[\pi2s_{1/2},\nu2s_{1/2}]^{J=1^+}$&     -1.379&       0.330&       2.443&       0.807 \\
	$[\pi1d_{3/2},\nu1d_{3/2}]^{J=1^+}$&     -0.726&      -0.095&      -1.548&       0.147 \\
	\hline
	$\sum b_{ph}$&&  &&   1.028\\
	$P(\nu)$&&  &&   1.056\\
	\midrule 
	\multicolumn{3}{c}{$2p-2h$ conf. }&$E_{2p-2h}$   &$N_{php'h'}$\\
	\multicolumn{3}{c}{$\big[[\pi1d_{5/2}\nu2s_{1/2}]_{J_H=3}[\pi2s_{1/2}\pi1d_{3/2}]_{J_P=2}\big]_{J_{T}=1}$} &   9.407 &  0.353  \\
	\multicolumn{3}{c}{$\big[[\pi1d_{5/2}\nu1d_{3/2}]_{J_H=2}[\pi1d_{3/2}\pi1d_{3/2}]_{J_P=2}\big]_{J_{T}=1}$} &  12.946 &  0.087  \\
	\multicolumn{3}{c}{$\big[[\pi1d_{5/2}\nu1d_{3/2}]_{J_H=1}[\pi1d_{3/2}\pi1d_{3/2}]_{J_P=0}\big]_{J_{T}=1}$} &  12.946 &  0.071  \\
	\multicolumn{3}{c}{$\big[[\pi1d_{5/2}\nu1d_{5/2}]_{J_H=1}[\pi2s_{1/2}\pi2s_{1/2}]_{J_P=0}\big]_{J_{T}=1}$} &   5.867 &  0.057  \\
	\midrule 
\end{tabular}
}
%
%
\resizebox{\columnwidth}{!}{%
\begin{tabular}{ccccc}
\toprule
&&E$_{SSRPA}$=      -0.364    MeV &&\\
\hline
$1p-1h$ conf. &$E_{1p-1h}$   &$N_{ph}$& &\\
$[\pi2s_{1/2},\nu2s_{1/2}]^{J=1^+}$&     -1.379&   0.017  &&\\
$[\pi1d_{3/2},\nu2s_{1/2}]^{J=1^+}$&      2.161&   0.012  &&\\
$[\pi1d_{3/2},\nu1d_{3/2}]^{J=1^+}$&     -0.726&   0.038  &&\\
&&&$N_1=       0.078$&$N_2=       0.922 $ \\
\hline
$1p-1h$ conf. &$E_{1p-1h}$   &$A_{ph}$& $F_{ph}$&$b_{ph}$\\
\hline
$[\pi1d_{3/2},\nu1d_{5/2}]^{J=1^+}$&      6.038&       0.061&       3.094&       0.188 \\
$[\pi2s_{1/2},\nu2s_{1/2}]^{J=1^+}$&     -1.379&      -0.129&       2.443&      -0.316 \\
$[\pi1d_{3/2},\nu1d_{3/2}]^{J=1^+}$&     -0.726&       0.195&      -1.548&      -0.302 \\
\hline
$\sum b_{ph}$&&  &&  -0.430\\
$P(\nu)$&&  &&   0.185\\
\midrule 
\multicolumn{3}{c}{$2p-2h$ conf. }&$E_{2p-2h}$   &$N_{php'h'}$\\
\multicolumn{3}{c}{$\big[[\pi1d_{5/2}\nu1d_{3/2}]_{J_H=1}[\pi1d_{3/2}\pi1d_{3/2}]_{J_P=2}\big]_{J_{T}=1}$} &  12.946 &  0.382  \\
\multicolumn{3}{c}{$\big[[\pi1d_{5/2}\nu1d_{3/2}]_{J_H=1}[\pi1d_{3/2}\pi1d_{3/2}]_{J_P=0}\big]_{J_{T}=1}$} &  12.946 &  0.063  \\
\multicolumn{3}{c}{$\big[[\pi1d_{5/2}\nu1d_{3/2}]_{J_H=3}[\pi1d_{3/2}\pi1d_{3/2}]_{J_P=2}\big]_{J_{T}=1}$} &  12.946 &  0.056  \\
\midrule 
\end{tabular}
}
\end{table}

\begin {table}
\caption {\label{tab-3}  As in Table \ref{tab-2} but for the states at 0.185 and 0.247 MeV.
}
\resizebox{\columnwidth}{!}{%
	\begin{tabular}{ccccc}
		\toprule
		&&E$_{SSRPA}$=       0.185    MeV &&\\
		\hline
		$1p-1h$ conf. &$E_{1p-1h}$   &$N_{ph}$& &\\
		$[\pi2s_{1/2},\nu2s_{1/2}]^{J=1^+}$&     -1.379&   0.027  &&\\
		$[\pi1d_{3/2},\nu2s_{1/2}]^{J=1^+}$&      2.161&   0.016  &&\\
		&&&$N_1=       0.046$&$N_2=       0.954 $ \\
		\hline
		$1p-1h$ conf. &$E_{1p-1h}$   &$A_{ph}$& $F_{ph}$&$b_{ph}$\\
		\hline
		$[\pi1d_{3/2},\nu1d_{5/2}]^{J=1^+}$&      6.038&       0.050&       3.094&       0.153 \\
		$[\pi2s_{1/2},\nu2s_{1/2}]^{J=1^+}$&     -1.379&       0.164&       2.443&       0.400 \\
		$[\pi1d_{3/2},\nu1d_{3/2}]^{J=1^+}$&     -0.726&       0.014&      -1.548&      -0.022 \\
		\hline
		$\sum b_{ph}$&&  &&   0.534\\
		$P(\nu)$&&  &&   0.285\\
		\midrule 
		\multicolumn{3}{c}{$2p-2h$ conf. }&$E_{2p-2h}$   &$N_{php'h'}$\\
		\multicolumn{3}{c}{$\big[[\pi1d_{5/2}\nu1d_{3/2}]_{J_H=3}[\pi1d_{3/2}\pi1d_{3/2}]_{J_P=2}\big]_{J_{T}=1}$} &  12.946 &  0.249  \\
		\multicolumn{3}{c}{$\big[[\pi1d_{5/2}\nu2s_{1/2}]_{J_H=2}[\pi2s_{1/2}\pi1d_{3/2}]_{J_P=2}\big]_{J_{T}=1}$} &   9.407 &  0.167  \\
		\multicolumn{3}{c}{$\big[[\pi1d_{5/2}\nu1d_{3/2}]_{J_H=1}[\pi1d_{3/2}\pi1d_{3/2}]_{J_P=0}\big]_{J_{T}=1}$} &  12.946 &  0.077  \\
		\multicolumn{3}{c}{$\big[[\pi1d_{5/2}\nu1d_{3/2}]_{J_H=3}[\pi2s_{1/2}\pi1d_{3/2}]_{J_P=2}\big]_{J_{T}=1}$} &   9.407 &  0.053  \\
		\midrule 
	\end{tabular}
}
%
%
\resizebox{\columnwidth}{!}{%
\begin{tabular}{ccccc}
	\toprule
	&&E$_{SSRPA}$=       0.247    MeV &&\\
	\hline
	$1p-1h$ conf. &$E_{1p-1h}$   &$N_{ph}$& &\\
	$[\pi2s_{1/2},\nu2s_{1/2}]^{J=1^+}$&     -1.379&   0.092  &&\\
	$[\pi2s_{1/2},\nu1d_{3/2}]^{J=1^+}$&     -4.266&   0.011  &&\\
	$[\pi1d_{3/2},\nu1d_{3/2}]^{J=1^+}$&     -0.726&   0.024  &&\\
	&&&$N_1=       0.131$&$N_2=       0.869 $ \\
	\hline
	$1p-1h$ conf. &$E_{1p-1h}$   &$A_{ph}$& $F_{ph}$&$b_{ph}$\\
	\hline
	$[\pi1d_{3/2},\nu1d_{5/2}]^{J=1^+}$&      6.038&      -0.013&       3.094&      -0.039 \\
	$[\pi2s_{1/2},\nu2s_{1/2}]^{J=1^+}$&     -1.379&       0.303&       2.443&       0.740 \\
	$[\pi1d_{3/2},\nu1d_{3/2}]^{J=1^+}$&     -0.726&      -0.154&      -1.548&       0.238 \\
	\hline
	$\sum b_{ph}$&&  &&   0.941\\
	$P(\nu)$&&  &&   0.885\\
	\midrule 
	\multicolumn{3}{c}{$2p-2h$ conf. }&$E_{2p-2h}$   &$N_{php'h'}$\\
	\multicolumn{3}{c}{$\big[[\pi1d_{5/2}\nu1d_{3/2}]_{J_H=3}[\pi1d_{3/2}\pi1d_{3/2}]_{J_P=2}\big]_{J_{T}=1}$} &  12.946 &  0.249  \\
	\multicolumn{3}{c}{$\big[[\pi1d_{5/2}\nu2s_{1/2}]_{J_H=2}[\pi2s_{1/2}\pi1d_{3/2}]_{J_P=2}\big]_{J_{T}=1}$} &   9.407 &  0.167  \\
	\multicolumn{3}{c}{$\big[[\pi1d_{5/2}\nu1d_{3/2}]_{J_H=1}[\pi1d_{3/2}\pi1d_{3/2}]_{J_P=0}\big]_{J_{T}=1}$} &  12.946 &  0.077  \\
	\multicolumn{3}{c}{$\big[[\pi1d_{5/2}\nu1d_{3/2}]_{J_H=3}[\pi2s_{1/2}\pi1d_{3/2}]_{J_P=2}\big]_{J_{T}=1}$} &   9.407 &  0.053  \\
	\midrule 
\end{tabular}
}
\end{table}

In Table \ref{tab-1}, we show the particle-hole configurations contributing the most to the norm (upper part) and to the GT transition probability (lower part) for the RPA  state located at 0.310 MeV and 1.422 MeV.  The superscripts $\pi$, $\nu$ identify  proton and neutron states, respectively, while $N_{ph}$ show the contribution of each $1p-1h$ configuration to the norm (see Eq. \ref{norm}), with unperturbed energy $E_{1p-1h}$, in MeV units. The state located at 1.422 MeV does not contribute to the $\beta$-decay lifetime, but we show it for the sake of completeness, being its energy close to the $\beta$-threshold. For the state at 0.310 MeV, the most important configurations  both to the norm and to the transition probability, are the $[\pi1d_{3/2},\nu1d_{3/2}]$ and $[\pi2s_{1/2},\nu2s_{1/2}]$, both of them having negative particle-hole excitation energy. A smaller contribution to the transition probability is also given
by the positive energy configuration $[\pi1d_{3/2},\nu1d_{5/2}]$ due to the large
corresponding matrix element of the transition operator $F_{ph}$. The same configurations appear to be important also for the state at 1.422 MeV. It is therefore possible that the increased strength observed in SSRPA is partially due to the shift toward lower energy of this state (or to its fragmentation) as a consequence of the coupling with the $2p-2h$ configurations. Indeed, in SSRPA, one can see that the number of states within the $\beta$-window significantly increases (see for example
Fig. \ref{Fig-Si34-SGII}), resulting in a lower lifetime compared to the RPA one. In Tables \ref{tab-2} and \ref{tab-3} , one can see the same analysis of Table \ref{tab-1} for the SSRPA states. The norm of these states is the sum of two contributions, corresponding to the $1p-1h$ ($2p-2h$) components, indicated as $N_1$ ($N_2$) in the table,  while $N_{php'h'}$ indicates the contribution of each $2p-2h$ configuration), with unperturbed energy $E_{2p-2h}$, in MeV units. Together with the $1p-1h$ components as in Table 1, we also show the $2p-2h$ most significant components, e.g. contributing more than 0.05 to the norm. The notation $\big[[hh']_{J_H}[pp']_{J_P}\big]_{J_{T}}$ is used, where $J_H(J_P)$ is the total angular momentum of the two hole (particle) states  and $J_T$ is the total angular momentum. One can see that all the states have a predominant $2p-2h$ nature (larger than 80 \%), which is fragmented over several configurations. Moreover, one can observe that the largest contributions involve the $\pi1d_{5/2},\nu1d_{3/2},\nu2s_{1/2}$ hole states, and the two particle levels $\pi1d_{3/2}$ and $\pi2s_{1/2}$. This is not surprising, as these levels are closest to the Fermi (neutron and proton) energies. At the same time, due to the quasi-boson approximation, the SSRPA transition probabilities are given by Eq. (\ref{prob}), e.g. they  are still driven by the $1p-1h$ configurations. One can see that the
$[\pi2s_{1/2},\nu2s_{1/2}]$ component contributes to all the states, both with respect to the $N_1$ content and to the transition probability itself. Also the $[\pi1d_{3/2},\nu1d_{3/2}]$ (and to a less extent the $[\pi1d_{3/2},\nu1d_{5/2}]$) configuration plays an important role in building up the collectivity of these states. Summarizing, one can see that the effect of the $2p-2h$ configurations introduced in the SSRPA is to increase the density of states within the $\beta$-window,  these states having a predominant $2p-2h$ nature, while the $1p-1h$ one is fragmented over the same kind of configurations which appear in the RPA states of Table \ref{tab-1}.
\section{Conclusions}
In this work, we study the effect of the inclusion of the $2p-2h$ configurations within the SSRPA on the $\beta$-decay half-lives of four closed shell nuclei, namely  $^{24}$O,  $^{34}$Si, $^{78}$Ni, and $^{132}$Sn. We employ four Skyrme forces and, as a general trend, we observe a significant improvement of the predicted $\beta$ half-lives with respect to the RPA ones, without the use of any \textit{ad hoc} quenching factor. The explicit inclusion of the $2p-2h$ configurations produces a more fragmented and richer spectrum within the $\beta$-window, resulting in lower $\beta$ half-lives.
As an illustrative case, the properties of the most collective states in $^{34}$Si are also discussed in terms of the most important $1p-1h$ configurations, contributing both to their norm and transition probability. Their $2p-2h$ nature, both in terms of the norm content and the most important configurations, is also discussed. We also studied the effect of the  $J^2$-terms for two different Skyrme interactions, e.g. SGII  and SLy5. We see that these terms  might have an important role, quantitatively depending both on the nucleus and force parametrization. However, their effect is comparable with those of the tensor interaction  and therefore  their interplay should be considered, especially in the derivation of new functionals for beyond-mean field approaches, targeting the description of the $\beta$-decay and GT excitations.
Moreover, more systematic studies, including open-shell nuclei, are necessary to fully assess the impact of beyond mean-field correlations introduced within the SSRPA approach. Work is currently underway to develop a Second Quasi-particle RPA approach, which can be numerically implemented for spherical nuclei.

\end{document}